\documentclass[12pt]{iopart}

\expandafter\let\csname equation*\endcsname\relax
\expandafter\let\csname endequation*\endcsname\relax 

\usepackage[english]{babel}
\usepackage{amsmath}
\usepackage{amssymb}
\usepackage{graphicx}
\usepackage{color}
\usepackage[utf8]{inputenc}
\usepackage{amsmath}
\usepackage{amsfonts}
\usepackage{graphicx} 
\usepackage{amsthm}
\usepackage{placeins}
\usepackage{verbatim}
\usepackage{booktabs}
\usepackage{color}
\usepackage{bm}
\usepackage{bbold}
\usepackage[colorlinks=true,linkcolor=red,citecolor=blue]{hyperref}
\usepackage{comment}
\usepackage{url}
\usepackage[all]{xy}
\usepackage[width=17.5cm, vscale=0.8]{geometry}
\newcommand{\beq}{\begin{equation}}
\newcommand{\eeq}{\end{equation}}
\newcommand{\be}{\begin{equation}}
\newcommand{\ee}{\end{equation}}

\DeclareMathOperator*{\intinf}{\int_{-\infty}^{+\infty}\!\!\!}

\begin{document}

\title{Index Distribution of Cauchy Random Matrices}

\author{Ricardo Marino, Satya N. Majumdar, Gr\'egory Schehr and Pierpaolo Vivo}

\address{Laboratoire de Physique Th\'eorique et Mod\`eles Statistiques, UMR 8626, Universit\'e Paris Sud 11 and CNRS, B\^at. 100, Orsay F-91405, France}

\date{\today}

\begin{abstract}
Using a Coulomb gas technique, we compute analytically the probability $\mathcal{P}_\beta^{(C)}(N_+,N)$ that a large $N\times N$ Cauchy random matrix has $N_+$ positive eigenvalues, where $N_+$ is called the \emph{index} of the ensemble. We show that this probability scales for large $N$ as $\mathcal{P}_\beta^{(C)}(N_+,N)\approx \exp\left[-\beta N^2 \psi_C(N_+/N)\right]$, where $\beta$ is the Dyson index of the ensemble. The rate function $\psi_C(\kappa)$ is computed in terms of single integrals that are easily evaluated numerically and amenable to an asymptotic analysis. We find that the rate function, around its minimum at $\kappa=1/2$, has a quadratic behavior modulated by a logarithmic singularity. As a consequence, the variance of the index scales for large $N$ as $\mathrm{Var}(N_+)\sim \sigma_C\ln N$, where $\sigma_C=2/(\beta\pi^2)$ is twice as large as the corresponding prefactor in the Gaussian and Wishart cases. The analytical results are checked by numerical simulations and against an exact finite $N$ formula which, for $\beta=2$, can be derived using orthogonal polynomials.

\end{abstract}

\maketitle


\section{Introduction}

Ensembles of matrices with random entries have been extensively studied since the seminal works of Wigner \cite{wigner_51}, Dyson \cite{dyson_62} and Mehta \cite{dyson_mehta_62}. However, many years before the official birth of Random Matrix Theory (RMT) in nuclear physics in the 1950s, statisticians had already introduced some of the RMT machinery in their studies on multivariate analysis \cite{wishart,fisher,hsu}. Restricting our scope to matrices with real spectra, two main classes of ensembles are typically considered, $\iota)$ matrices with independent entries, and $\iota\iota)$ matrices with rotational invariance. While limited analytical insight is generally available for the former, rotationally invariant ensembles of $N\times N$ matrices are generally characterized by a joint probability density function (jpdf) of the $N$ real eigenvalues of the form

\be
P_\beta({\bm \lambda})=\frac{1}{Z_{N,\beta}}\prod_{j<k}|\lambda_j-\lambda_k|^\beta \prod_{j=1}^N e^{-\beta V(\lambda_j)} \;, \label{jpdgeneral}
\ee
where $Z_{N,\beta}$ is a normalization constant, and $\beta=1,2,4$ is the Dyson index of the ensemble, identifying real symmetric, complex Hermitian and quaternion self-dual matrices, respectively. $V(x)$, the \emph{potential}, is a function suitably growing at infinity that defines the model. For instance, $V(x)=x^2/2$ for the Gaussian ensemble, or $V(x)=x-\alpha \ln x$ for the Wishart-Laguerre ensemble.

Armed with \eqref{jpdgeneral}, a wealth of statistical questions about eigenvalue distributions can be efficiently tackled for both finite and large $N$, such as the average density of states, gap distributions and statistics of extreme eigenvalues. While usually the focus is on \emph{typical} fluctuations of such random variables, several interesting cases have been lately considered, where the interest lies instead on \emph{atypical} (rare) fluctuations, far away from the average (see Ref. \cite{majumdar_schehr_13} for a review). To mention just a few, the large deviation probability of extreme eigenvalues of Gaussian~\cite{dean_majumdar_08,majumdar_vergassola_09,borot_eynard_majumdar_nadal_10,saito_10,benArous_dembo_guionnet_01} and Wishart random matrices~\cite{majumdar_vergassola_09,vivo_majumdar_bohigas_07,katzav_perezCastillo_10}, the number of stationary points of random Gaussian landscapes~\cite{bray_dean_07,fyodorov_williams_07,fyodorov_nadal_12}, the distribution of free energies in mean-field spin glass models~\cite{parisi_rizzo_09,monthus_garel_10}, conductance and shot noise distributions in chaotic mesoscopic cavities~\cite{vivo_majumdar_bohigas_07,vivo_majumdar_bohigas_10}, entanglement entropies of a pure random state of a bipartite quantum system~\cite{facchi_marzolino_parisi_pascazio_scardicchio_08,nadal_majumdar_vivo_10,pasquale_facchi_parisi_pascazio_09,vivo_11} and the mutual information in multiple input multiple output (MIMO) channels~\cite{moustakas_11}. In addition, RMT has also proven an invaluable tool in understanding large deviation properties of various observables in the so called vicious walker (or nonintersecting Brownian motion) problem~\cite{schehr_majumdar_comtet_randonFurling_08,nadal_majumdar_09,rambeau_schehr_10,forrester_majumdar_schehr_11,schehr_majumdar_comtet_forrester_12}. A powerful tool to deal with such instances is the Coulomb gas technique, originally popularized\footnote{E.P. Wigner had however used it already in 1957 \cite{wignercanada} to compute the density of states of the Gaussian ensemble.} by Dyson~\cite{dyson_62}, which will be reviewed in Section \ref{coulomb_gas}.

Perhaps the simplest and most natural of such questions concerns the random variable $N_{\mathcal{I}}$, defined as the number of eigenvalues contained in an interval $\mathcal{I}$ on the real line. The average value $\langle N_{\mathcal{I}}\rangle$ can be clearly computed as $\langle N_{\mathcal{I}}\rangle=N\int_{\mathcal{I}}dx \rho(x)$, where $\rho(x)$ is the average density of eigenvalues of the ensemble. What about its fluctuations? Dyson ~\cite{dyson_62} studied the variance of the number of eigenvalues in the ``bulk limit", i.e. when $\mathcal{I}=[-\delta_N L/2,\delta_N L/2]$, where $\delta_N=\pi/\sqrt{2N}$ is the mean spacing in the bulk and $L$ is kept fixed as $N\to\infty$. Clearly $\langle N_{\mathcal{I}}\rangle=L$ and the variance grows logarithmically with $L$,
\begin{equation}
\langle(N_\mathcal{I}-L)^2\rangle\sim \frac{2}{\beta\pi^2}\ln(L)+B_\beta\label{dysonbulk} \;,
\end{equation}
where the constant $B_\beta$ was computed by Dyson and Mehta \cite{dyson_mehta_62}. Therefore the typical scale of fluctuations around the mean is $\sqrt{\ln(L)}$, and the computation of higher moments \cite{CL,FS} reveals that on this scale, the random variable $N_{\mathcal{I}}$ has a Gaussian distribution.

Another related observable, which on the contrary has surprisingly escaped a thorough investigation until very recently, is the \emph{index} $N_+(\zeta)=\sum_{i=1}^N\theta(\lambda_i-\zeta)$, defined as the number of eigenvalues exceeding a threshold $\zeta$. In particular, we will focus on the number $N_+:=N_+(0)$ of positive eigenvalues. Note that in this case, where $\mathcal{I}$ is the full unbounded interval $[0,\infty)$ the previous result \eqref{dysonbulk}, valid on a small symmetric interval around the origin, is no longer applicable.

This random variable $N_+$ naturally arises in the study of the stability of a multidimensional potential landscape $V(x_1,x_2,\ldots,x_N)$~\cite{wales}. For instance, in string theory $V$ may represent the potential associated with a moduli space~\cite{douglas_03}. As far as glassy systems are concerned, the point $\{x_i\}$ may instead represent a configuration of the system and $V(\{x_i\})$ the energy of that configuration~\cite{cavagna_garrahan_giardina_00}. Generally, for disordered systems $V(\{x_i\})$ may represent the free energy landscape. Typically this $N$-dimensional landscape displays a complex pattern of stationary points that play a relevant role both in statics and dynamics of such systems~\cite{wales}. The stability of a stationary point of this $N$-dimensional landscape depends on the $N$ real eigenvalues of the $(N\times N)$ Hessian matrix $M_{i,j}= \left[\partial^2 V/{\partial x_i \partial x_j}\right]$ which is symmetric. If all the eigenvalues are positive (negative), the stationary point is a local minimum (local maximum). If some, but not all, are positive then the stationary point is a saddle. The number of positive eigenvalues (the index), $0\le  N_{+} \le N$, is therefore a crucial indicator of how many directions departing from the stationary point are stable. Given a random potential $V$, the entries of the Hessian matrix at a stationary point are usually correlated. However, often important insights can be obtained by discarding these correlations. In the simplest case, one may assume that the entries of the Hessian matrix are independent Gaussian variables. This then leads to the index problem for a real symmetric Gaussian matrix. This toy model, called the random Hessian model (RHM), has been studied extensively in the context of disordered systems~\cite{cavagna_garrahan_giardina_00}, landscape based string theory~\cite{aazami_easther_06}, quantum cosmology~\cite{mersiniHoughton_05}, random supergravity theories \cite{supergravity} and multifield inflation theories \cite{multifield}. 

For Gaussian matrices with $\beta=1$, the statistics of $N_+$ was considered by Cavagna \textit{et al.} \cite{cavagna_garrahan_giardina_00}. Using replica methods with Grassmann variables, they found that around its mean value $N/2$, the random variable $N_+$ has typical fluctuations of $\mathcal{O}(\sqrt{\ln N})$ for large $N$. Moreover, the distribution of these typical fluctuations is Gaussian, i.e.
\begin{equation}
\mathcal{P}^{(G)}_\beta (N_+,N)\approx\exp\left[-\frac{\pi^2}{2 \ln (N)}(N_+-N/2)^2\right]\label{Gausscavagna} \;,
\end{equation}
where this form of the distribution is valid over a region of $\mathcal{O}(\sqrt{\ln N})$ around the mean $\langle N_+\rangle=N/2$ for large $N$. This implies that  variance grows logarithmically with $N$, $\mathrm{Var}(N_+)\sim \sigma_G\ln N$, with $\sigma_G=1/\pi^2$, for large $N$. For atypically large fluctuations ($N_+\gg N/2$), the Gaussian distribution \eqref{Gausscavagna} is no longer valid, and the large deviation tails were computed in \cite{majumdar_nadal_scardicchio_vivo_09,majumdar_nadal_scardicchio_vivo_11}, this time for all $\beta$ using a Coulomb gas method. Setting $\kappa=N_+/N$, the probability density of the random variable $N_+$ was found to scale for large $N$ as\footnote{Hereafter, $\approx$ stands for a logarithmic equivalence, $\lim_{N\to\infty}\frac{-\ln \mathcal{P}^{(G)}_\beta (N_+=\kappa N,N)}{\beta N^2}=\psi_G(\kappa)$.}

\be
\mathcal{P}^{(G)}_\beta (N_+=\kappa N,N)\approx \exp\left[-\beta N^2 \psi_G(\kappa)\right]\label{largedevgauss}  \;,
\ee
where the rate function $\psi_G(\kappa)$ was computed exactly \cite{majumdar_nadal_scardicchio_vivo_09,majumdar_nadal_scardicchio_vivo_11} over the full range $0\leq \kappa\leq 1$. It is independent of $\beta$ and has a minimum at $\kappa=1/2$ (corresponding to the typical situation, where $\langle N_+\rangle=N/2$ i.e. half of the eigenvalues are positive on average). The case $\kappa=1$ (and similarly $\kappa=0$) corresponds to the extreme situation where \emph{all} eigenvalues are positive (negative). Therefore $\mathcal{P}^{(G)}_\beta (N_+=N,N)=\mathrm{Prob}[\lambda_1>0,\ldots,\lambda_N>0]=\mathrm{Prob}[\lambda_{\mathrm{min}}>0]$, i.e. the probability that the smallest eigenvalue $\lambda_{\mathrm{min}}$ is positive. Hence there is a natural connection between the index problem and the distribution of extreme eigenvalues, tackled in \cite{dean_majumdar_08,majumdar_vergassola_09,vivo_majumdar_bohigas_07,dean_majumdar_06,ramli_katzav_perezCastillo_12}. Note that the index problem in the complex plane (i.e. the statistics of the number of complex eigenvalues with modulus greater than a threshold) has also been recently considered \cite{allez_touboul_wainrib_13,armstrong_serfaty_zeitouni_13}.

Expanding the rate function $\psi_G(\kappa)$ around the minimum, it was found that it does not display a simple quadratic behavior as one could have naively expected. Instead, the quadratic behavior is modulated by a logarithmic singularity, implying that in the close vicinity of $N_+=N/2$ over a scale of $\sqrt{\ln N}$ one recovers a Gaussian distribution,
\be
\mathcal{P}^{(G)}_\beta \left(N_+,N\right)\approx \exp\left[-\frac{\left(N_+-N/2\right)^2}{2\left(\mathrm{Var}(N_+)\right)}\right]\mbox{ for }N_+\to N/2\label{Pintrovargauss} \;,
\ee
with

\be
\mathrm{Var}(N_+)\sim \sigma_G\ln N + C_\beta+\mathcal{O}(1/N) \;, \; \sigma_G = \frac{1}{\beta \pi^2} \label{varlargeN} \;.
\ee
Note that for $\beta=1$ one recovers the result in \cite{cavagna_garrahan_giardina_00}. The constant term $C_2$ was found \cite{majumdar_nadal_scardicchio_vivo_11} via the asymptotic expansion of a finite-$N$ variance formula conjectured by Prellberg and later rigorously established \cite{witte_forrester_12}. Interestingly, the same analysis performed on positive definite Wishart matrices \cite{majumdar_vivo_12} for a threshold at $\zeta$ within the support of the spectral density leads to 
\be
\mathrm{Var}(N_+(\zeta))\sim \sigma_G\ln N +\mathcal{O}(1)\label{varlargeNwishart} \;,
\ee
where the leading term is independent of $\zeta$ and exactly identical to the Gaussian case. Note that an explicit formula for the full probability of the index for \emph{finite} $N$ is not available to date in either case.

Given the rather robust large $N$ behavior of the variance which holds both for Gaussian matrices (\ref{varlargeN}) and Wishart matrices (\ref{varlargeNwishart}), we investigate here the index probability distribution of yet another ensemble of random matrices for which we expect a 
different behavior. We consider indeed the Cauchy ensemble of $N\times N$ matrices $\mathbf{H}$ which are real symmetric ($\beta=1$), complex Hermitian ($\beta=2$) or quaternion self-dual ($\beta=4$). The Cauchy ensemble is characterized by the following probability measure
\be
P(\mathbf{H})\propto \left[\det\left({\bm 1}_N+{\mathbf{H}}^2\right)\right]^{-\beta(N-1)/2-1} \;, \label{defcauchy}
\ee
where ${\bm 1}_N$ is the identity matrix $N\times N$. The definition \eqref{defcauchy} is evidently invariant under the similarity transformation $\mathbf{H}\to \mathbf{U H U}^{-1}$, where $\mathbf{U}$ is an orthogonal $(\beta=1)$, unitary $(\beta=2)$ or symplectic $(\beta=4)$ matrix. The jpdf of the $N$ real eigenvalues can be then immediately written as
\be
P_\beta({\bm \lambda})\propto \prod_{j=1}^N \frac{1}{(1+\lambda_j^2)^{\beta(N-1)/2+1}}\prod_{i<k}|\lambda_i-\lambda_k|^\beta \;.\label{jpdcauchy}
\ee
As in the Gaussian and Wishart cases, we can give an electrostatic interpretation of the jpdf \eqref{jpdcauchy}, where the $\lambda_i$s are positions of charged particles (with say positive unit charge) on the real line and repelling each other via the 2d-Coulomb (logarithmic) interaction. Here they feel an additional interaction with a single particle, with charge $-(N-1+2/\beta)$ placed at the point of coordinate $(0,1)$ in the 2d plane. A closely related ensemble occurs in the context of mesoscopic transport where it represents the scattering matrix of a quantum dot coupled to the outside world by non ideal leads containing $N$ scattering channels \cite{brouwer_95,fyodnock}. It is also one of the typical examples where free probability theory efficiently applies in the context of random matrices models
\cite{Burda02,Burda11}. Interestingly, the Cauchy ensemble (\ref{jpdcauchy}) is related to the circular ensemble of RMT via the stereographic projection \cite{forrester}. Indeed, if one defines the angles $\theta_k$ via the relation
\be
e^{\mathrm{i} \theta_k} = \frac{1-\mathrm{i} \lambda_k}{1+\mathrm{i} \lambda_k} \;,
\ee
then the jpdf of the $\theta_k$'s is precisely the one of the eigenvalues in the $\beta$-circular ensemble. This implies in particular that local fluctuations of the eigenvalues in the Cauchy ensemble are described, for large $N$, by the sine-kernel \cite{BO01}. This connection with the circular  ensembles implies also that, in contrast to most other random matrix models, the finite-$N$ spectral density $\rho_N(\lambda)$ is independent of $N$, i.e. it coincides for all $N$ with its asymptotic expression $\rho^\star_C(\lambda)$. This density has fat tails extending over the full real axis, and its expression is given for all $\beta$ by \cite{forrester,tierz_01}
\be
\rho^\star_C(\lambda)=\frac{1}{\pi}\frac{1}{1+\lambda^2} \;.
\label{rho_cauchy_standard}
\ee
We will also see below that the Cauchy ensemble, for $\beta = 2$, possesses the remarkable property of being exactly solvable for finite $N$ and $\beta=2$, as the suitable orthogonal polynomials can be determined in terms of Jacobi polynomials (see Section \ref{exact_N} for details).
Hence, a major difference with Gaussian or Wishart matrices for which the mean spectral density has a finite support is that in the case of Cauchy matrices, the density $\rho^\star_C(\lambda)$ has no edge (\ref{rho_cauchy_standard}). It was recently shown in \cite{majumdar_schehr_villamaina_vivo_13} that, for large $N$, the absence of an edge has indeed important consequences on the right large deviations of the top eigenvalue, $\lambda_{\max} = \max_{1 \leq i \leq N} \lambda_i$, in this ensemble (see also Refs. \cite{witte_forrester_00,NNR09} for the study of $\lambda_{\max}$ for $\beta = 2$, though the large deviations were not studied there). 

The purpose of this paper is to study the full probability distribution of the index for the Cauchy ensemble, using a Coulomb gas technique. As a bonus, we also obtain the \emph{constrained} spectral density of a Cauchy ensemble with a prescribed fraction of positive eigenvalues. In the limit $\kappa\to 0$ (where all eigenvalues are negative), we recover the large deviation law for the largest Cauchy eigenvalue derived in \cite{majumdar_schehr_villamaina_vivo_13}.
We show that the probability distribution that a Cauchy matrix has a fraction $\kappa=N_+/N$ of positive eigenvalues decays for large $N$ as
\be
\mathcal{P}^{(C)}_\beta (N_+=\kappa N,N)\approx \exp\left[-\beta N^2 \psi_C(\kappa)\right]\label{largedevgauss_2} \;,
\ee
where the rate function $\psi_C(\kappa)$, defined for $0\leq \kappa\leq 1$, is independent of $\beta$ and is calculated exactly (in terms of single integrals that cannot be further simplified) in \eqref{rate1} and \eqref{actionI}. The rate function has the following behavior close to the minimum at $\kappa=1/2$,

\be
\psi_C(\kappa=1/2+\delta)\sim  -\frac{\pi^2}{2}\frac{\delta^2}{\ln |\delta|} \mbox{ for }\delta\to 0,
\ee
resulting in a Gaussian distribution of the index around the typical value $\langle N_+\rangle=N/2$, albeit with a variance growing with $\ln N$, i.e.

\be
\mathcal{P}^{(C)}_\beta \left(N_+,N\right)\approx \exp\left[-\frac{\left(N_+-N/2\right)^2}{2\left(\mathrm{Var}(N_+)\right)}\right]\mbox{ for }N_+\to N/2 \;, \label{Pintrovar}
\ee
where
\be
 \mathrm{Var}(N_+)\sim \sigma_C \ln N +\mathcal{O}(1) \;, \; \sigma_C = \frac{2}{\beta \pi^2} = 2 \sigma_G \label{variance_cauchy} \;.
\ee
This result (\ref{variance_cauchy}) obtained via a Coulomb gas approach, valid for any $\beta$, is confirmed by an exact finite-$N$ formula, using orthogonal polynomials, for $\beta=2$. An interesting feature of this result (\ref{variance_cauchy}) is that the prefactor of the leading behavior $\propto \ln N$ of the variance is twice as large here as in the Gaussian \eqref{varlargeN} and Wishart  \eqref{varlargeNwishart} cases: $\sigma_C = 2 \sigma_ G$. Given that the local bulk statistics in all these 
cases is described by the sine-kernel, one may argue that this factor of $2$ is due to the presence of an edge in the density of eigenvalues for Gaussian and Wishart matrices, which is absent for Cauchy matrices. A thorough investigation of this issue is deferred to a separate publication \cite{MMSVprep}.

The plan of the paper is as follows. In section \ref{coulomb_gas}, we introduce the Coulomb gas formulation of the index problem, in terms of the minimization of an action which leads to a singular integral equation for the constrained density of eigenvalues. This integral equation is solved, in section \ref{coulomb_gas}, using the resolvent method. In section \ref{computation_of_rate} we present the evaluation of the action at the constrained density, from which we compute the rate function $\psi_C(\kappa)$ \eqref{largedevgauss_2} in terms of single integrals. Next, in section \ref{asymptotic}, we evaluate the asymptotic behavior of $\mathcal{P}^{(C)}_\beta \left(N_+,N\right)$ when $N_+$ is close to its average value ${N}/{2}$, extracting the leading behavior of the variance of $N_+$ as a function of $N$. Finally, in section \ref{exact_N}, we derive an exact finite $N$ formula for the variance of $N_+$ in the case $\beta=2$, showing a perfect agreement with the leading trend for large $N$, before concluding in section \ref{conclusion}. Some technical details have been confined to \ref{appendix_asympt} and \ref{exact_N_der}.


\section{Coulomb gas formulation and integral equation for the constrained density}
\label{coulomb_gas}

Let $N_+(\zeta)$ the number of eigenvalues of a Cauchy random matrix larger than $\zeta$. The probability density of $N_+(\zeta)$ is by definition
\be
\mathcal{P}_\beta^{(C)}(N_+(\zeta),N)=\int d^N\lambda P_\beta({\bm \lambda})\delta\left(N_+(\zeta)-\sum_{j=1}^N\theta(\lambda_j-\zeta)\right),\label{defint}
\ee
where $\theta$ is the Heaviside step function and $P_\beta({\bm \lambda})$ is defined in \eqref{jpdcauchy}.

We start by writing the jpdf $P_\beta({\bm \lambda})$ in exponential form, 
\be
P_\beta({\bm \lambda})\propto e^{-\beta H[{\bm \lambda}]},
\label{hamilt_cauchy}
\ee
where:
\be
H[{\bm \lambda}]=\left(\frac{N-1}{2}+\frac{1}{\beta}\right)\sum_{j=1}^N\ln(1+\lambda^2_j)-\sum_{i>j}\ln|\lambda_i-\lambda_j|.\label{hamiltonian}
\ee
In this form, the jpdf \eqref{hamilt_cauchy} mimics the Gibbs-Boltzmann weight of a system of charged particles in equilibrium under competing interactions. Following Dyson \cite{dyson_62} we can treat this system for large $N$ as a continuous fluid, described by a normalized density $\rho(\lambda)=(1/N)\sum_{i=1}^N\delta(\lambda-\lambda_i)$. Consequently, the multiple integral in \eqref{defint} can be converted into a functional integral in the space of normalizable densities. This procedure was first successfully employed to compute the large deviation of maximal eigenvalue of Gaussian matrices \cite{dean_majumdar_06} and afterwards applied in several different contexts \cite{majumdar_schehr_13,dean_majumdar_08,majumdar_vergassola_09,vivo_majumdar_bohigas_07,majumdar_nadal_scardicchio_vivo_11}.

In the continuum limit, the multiple integral \eqref{defint} becomes 
\be
\mathcal{P}_\beta^{(C)}(N_+(\zeta),N)\propto\int \mathcal{D}[\rho]\int dA_1\int dA_2\ e^{-\frac{\beta}{2}N^2 S[\rho]},\label{functional}
\ee
where the \emph{action} $S$ is given by
\begin{align}
S[\rho]=&\intinf dx\rho(x)\ln(1+x^2) - \int\intinf dxdx' \rho(x)\rho(x')\ln|x-x'| \\
&+ A_1\left(\intinf dx\rho(x)-1\right)+A_2\left(\int_\zeta^{+\infty}dx \rho(x)-\kappa\right) \;,
\label{eq:hamilt_1}
\end{align}
and $A_1,A_2$ are Lagrange multipliers, introduced to enforce the overall normalization of the density, and a fraction $\kappa$ of eigenvalues exceeding $\zeta$. 
The action $S$ has an evident physical meaning: it represents the free energy of the Coulomb fluid, whose particles are constrained 
to split over two regions of the real line, a fraction $\kappa$ to the right of $\zeta$ and a fraction $1-\kappa$ to the left of $\zeta$.
This free energy scales as $N^2$ (and not just $N$) because of the strong all-to-all interactions among the particles. The energetic component $\sim\mathcal{O}(N^2)$
of this free energy dominates over the entropic part $\sim\mathcal{O}(N)$, making it possible to use a saddle point method (see below). Note that the entropic
term has been thoroughly studied and employed to define interpolating ensembles in \cite{allez_bouchaud_guionnet_12,allez_bouchaud_majumdar_vivo_13}.

\begin{figure}[!htbp]
\centering
{
	\setlength{\fboxsep}{2pt}
	\fbox{\includegraphics[width=0.8\textwidth]{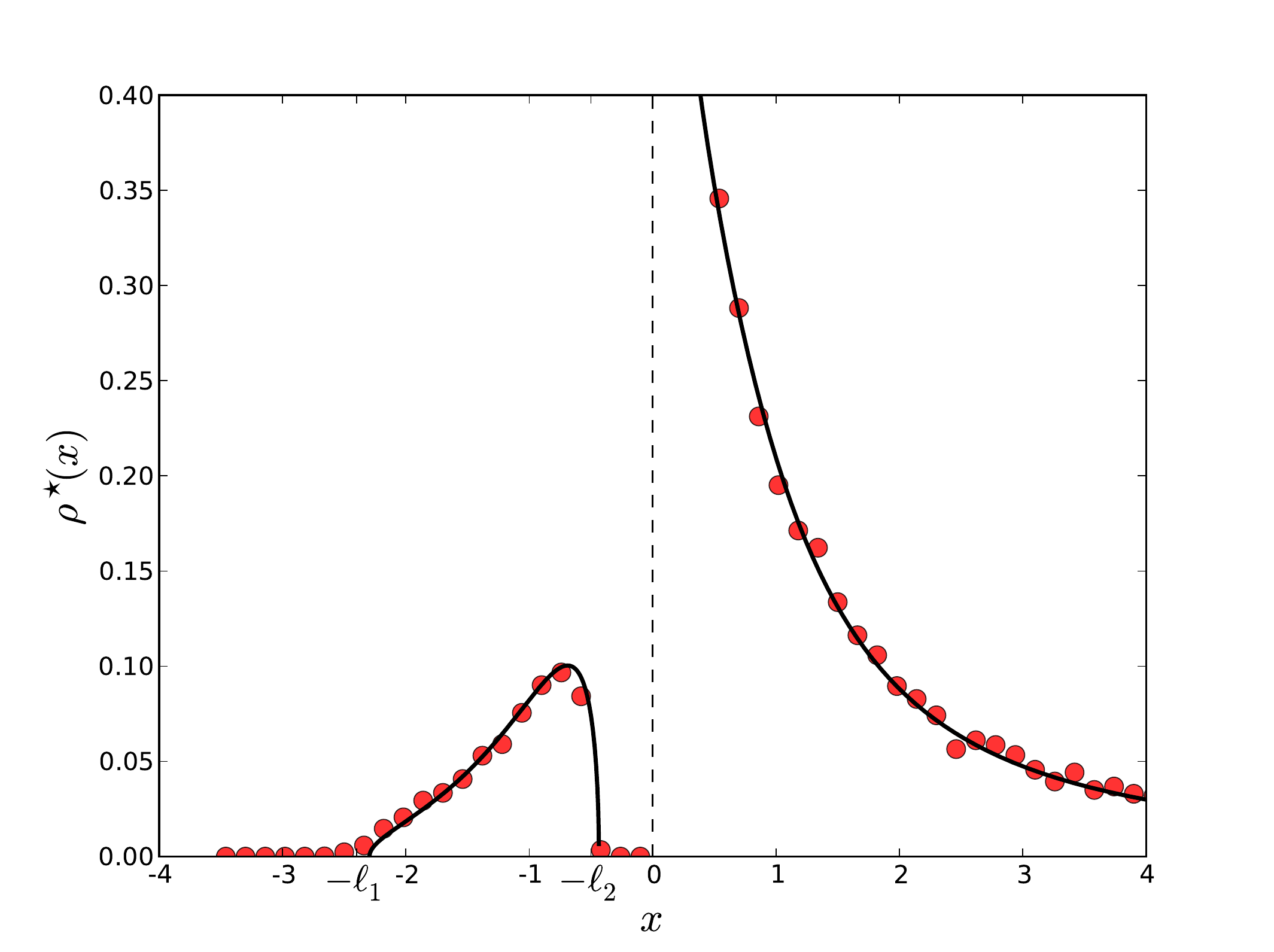}}
}
\caption{Constrained density of eigenvalues for a matrix of size $N=300$ where 90\% of the eigenvalues are forced to lie on the positive semi axis and the corresponding expected theoretical curve for $\kappa=0.9$ [$B=0.36613...$ from \eqref{intconstraint}].}
\label{fig:MC_with_function}
\end{figure}

As mentioned earlier, the integral \eqref{functional}, where we neglected terms of $\mathcal{O}(N)$, can be evaluated using a saddle point method for large $N$. The \emph{constrained} density of eigenvalues $\rho^\star(x)$ (depending parametrically on $\kappa$ and $\zeta$) is determined by the following variational condition

\be
\frac{\delta S}{\delta \rho}= \ln(1+x^2) -  2\int_{-\infty}^{+\infty}dx' \rho^\star(x')\ln|x-x'| + A_1+ A_2\theta(x-\zeta) =0 \;.\label{intlog}
\ee
This Eq. \eqref{intlog}, valid for $x$ inside the support of $\rho^\star(x)$, can be differentiated once with respect to $x$ to give the following singular integral equation
\be
\frac{x}{1+x^2} +A_2\delta(x-\zeta)= \mathcal{P}\int_{-\infty}^\infty \frac{\rho^\star(y)}{x-y}dy \;,
\label{eq:cauchy}
\ee
where $\mathcal{P}$ stands for Cauchy principal part. Solving \eqref{eq:cauchy} with the constraint $\int_\zeta^\infty dx\rho^\star(x)=\kappa$ is the main technical challenge. The physical intuition, supported by numerical simulations, points towards a density supported on two disconnected intervals (see Fig. \ref{fig:MC_with_function}): for $\kappa>1/2$, a compact blob with two edges to the left of $\zeta$, and a non-compact blob to the right of $\zeta$, extending all the way to infinity and with a singularity for $x\to\zeta^+$ (for $\kappa<1/2$ the situation is reversed, while only for $\kappa=1/2$ the two blobs merge into the unconstrained single-support density $\rho^\star_C(\lambda)$ \eqref{rho_cauchy_standard}). In such a situation, the standard inversion formula \cite{tricomi} for singular integral equation of the type in \eqref{eq:cauchy} cannot be applied, as it holds only for single support (``one-cut") solutions. However, a more general method, which we now present, allows to compute the \emph{resolvent} (or Green's function) for this two-cuts problem\footnote{See \cite{akemann} for a more sophisticated approach based on loop equation techniques.}, and from it to deduce $\rho^\star$.

We introduce the \emph{resolvent}
\be
G(z)=\int \frac{\rho^\star(x)}{z-x}dx \;, \label{defG}
\ee
for the Cauchy case. It is an analytic function in the complex plane outside the support of the density. From the resolvent, the density can be computed in the standard way as

\be
\frac{1}{\pi}\lim_{\varepsilon\to 0}\text{Im }G(x+\mathrm{i}\varepsilon)=\rho^\star(x) \;,\label{cauchysok}
\ee
where $\text{Im}$ stands for the imaginary part.

\underline{\textit{Unconstrained case.}} As a warm-up exercise, we first derive the resolvent equation (using purely algebraic manipulations) for the \emph{unconstrained} case (corresponding to \eqref{eq:cauchy} when $A_2=0$), where we expect to recover the density $\rho^\star_C(x)$ in Eq. \eqref{rho_cauchy_standard}.
First, we multiply both sides in \eqref{eq:cauchy} (dropping the principal value) by ${\rho^\star(x)}/{(z-x)}$ and we integrate it over $x$, obtaining
\be
\int \frac{x}{1+x^2}\frac{\rho^\star(x)}{z-x} dx = \iint\frac{\rho^\star(x)\rho^\star(y)}{(x-y)(z-x)}dxdy \;.\label{noprincipal}
\ee
Our goal is to express both sides in terms of $G(z)$ and, by doing so, obtain an algebraic equation for $G(z)$. We start by the right hand side (RHS) where we use the simple identity
\be
\frac{1}{(z-x)(x-y)} =\left(\frac{1}{z-x}+\frac{1}{x-y}\right)\frac{1}{z-y}\label{trick} \;.
\ee
Hence the RHS of \eqref{noprincipal} can be written as
\begin{align}
\nonumber\iint \frac{\rho^\star(x)\rho^\star(y)}{(x-y)(z-x)}dxdy=&\iint \left(\frac{1}{z-x}+\frac{1}{x-y}\right)\frac{\rho^\star(x)\rho^\star(y)}{z-y}dxdy\\ =& \iint \frac{\rho^\star(x)\rho^\star(y)}{(z-x)(z-y)}dxdy+\iint \frac{\rho^\star(x)\rho^\star(y)}{(x-y)(z-y)}dxdy \label{eq:RHS_2}.
\end{align}
The second term of the sum in \eqref{eq:RHS_2} (under the replacement $x\leftrightarrow y$) is the original RHS of \eqref{noprincipal} with the sign changed. The first term of the sum is just $G(z)^2$. Hence, we have that the RHS of \eqref{noprincipal} is just equal to $(1/2)G(z)^2$:
\be\label{rhs_gzsq}
\iint\frac{\rho^\star(x)\rho^\star(y)}{(x-y)(z-x)}dxdy = \frac{1}{2} G(z)^2 \;.
\ee

The left hand side (LHS) of \eqref{noprincipal} requires a little more algebraic manipulation to be expressed in terms of $G(z)$. We manipulate this expression in two different ways and exploit the equality between the results to get rid of one integral. First, using the identity $x/(1+x^2) = 1/x - 1/(x(1+x^2))$ one has
\begin{align}
\int \frac{x}{1+x^2}\frac{\rho^\star(x)}{z-x}dx= \int \frac{1}{x}\frac{\rho^\star(x)}{z-x}dx-\int \frac{1}{x(1+x^2)}\frac{\rho^\star(x)}{z-x}dx \;.\label{eq:LHS_2}
\end{align}
Note that in this splitting the two integrals in the RHS are individually divergent due to the pole at $x=0$, but the divergence cancels out between the two. We may regularize each individually divergent integral by adding a small imaginary part in the denominator that is removed at the end of the calculation. Using the relation \eqref{trick}, we may express the first term of the sum in \eqref{eq:LHS_2} as
\begin{align}\label{trick2}
\int \frac{1}{x}\frac{\rho^\star(x)}{z-x}dx=&\int \left(\frac{1}{x}+\frac{1}{z-x}  \right)\frac{\rho^\star(x)}{z}dx=\frac{1}{z}\underbrace{\int \frac{\rho^\star(x)}{x}dx}_{a_0}+\frac{1}{z}\underbrace{\int \frac{\rho^\star(x)}{z-x}dx}_{G(z)}=\frac{a_0}{z}+\frac{G(z)}{z}.
\end{align}

The second term of the sum in \eqref{eq:LHS_2} will not be calculated for now, and will be called $-\alpha(z)$. Using this manipulation \eqref{trick2}, we have:
\be
\int \frac{x}{1+x^2}\frac{\rho^\star(x)}{z-x} dx = \frac{a_0}{z}+\frac{G(z)}{z} - \alpha(z).
\ee
Now, we use a different strategy, using the identity $x/(1+x^2) = (x-z)/(1+x^2) + z/(1+x^2)$, to obtain 
\begin{align}
\nonumber\int \frac{x}{1+x^2}\frac{\rho^\star(x)}{z-x}dx=&\int \frac{x+z-z}{1+x^2}\frac{\rho^\star(x)}{z-x}dx\\
=&-\underbrace{\int \frac{\rho^\star(x)}{1+x^2}dx}_{a_1}+z\int \frac{1}{1+x^2}\frac{\rho^\star(x)}{z-x}dx\label{eq:LHS_3}.
\end{align}
The first term in the sum \eqref{eq:LHS_3} is a constant, which we call $a_1$. Now we proceed to manipulate the second term in (\ref{eq:LHS_3}) to obtain
\begin{align}
z\int \frac{1}{1+x^2}\frac{\rho^\star(x)}{z-x}dx =& z\int \frac{x-z+z}{x(1+x^2)}\frac{\rho^\star(x)}{z-x}dx\\
=&-z\underbrace{\int \frac{\rho^\star(x)}{x(1+x^2)}dx}_{a_2} + z^2 \underbrace{\int\frac{1}{x(1+x^2)}\frac{\rho^\star(x)}{z-x}dx}_{\alpha(z)}\label{eq:LHS_4}.
\end{align}

Therefore, the LHS of \eqref{noprincipal} can also be written as $-a_1 - za_2 + z^2\alpha(z)$. We have then two distinct ways of writing the LHS, and we can use them both to cancel $\alpha(z)$:
\be
\frac{a_0}{z}+\frac{G(z)}{z} - \alpha(z)=\text{(RHS)}=
-a_1 - za_2 + z^2\alpha(z) \;.
\label{eq:sum_equations}
\ee
Eliminating $\alpha(z)$ and recalling that the RHS is equal to $(1/2)G(z)^2$ \eqref{rhs_gzsq}, we find the algebraic equation for $G(z)$
\be
 -a_1 + z(a_0-a_2) + zG(z) = \frac{(1+z^2)}{2}G(z)^2 \;.
\label{eq:algeb_1}
\ee

We proceed to determine the constants $a_0$, $a_1$ and $a_2$ using the normalization condition of the density $\rho^\star(x)$. From \eqref{defG} (setting $|z|\to\infty$), it implies that $G(z)$ should asymptotically go as $G(z)\sim 1/z$. Taking the limit $|z|\to\infty$ in equation \eqref{eq:algeb_1}, we find equations for the coefficients
\be
-a_1 + z(a_0-a_2) + z\left(\frac{1}{z}+\mathcal{O}(z^{-2})\right) = \frac{z^2}{2}\left(\frac{1}{z}+\mathcal{O}(z^{-2})\right)^2 \;,
\ee
which implies $a_0=a_2$ and $a_1=\frac{1}{2}$. Our algebraic equation, finally, becomes:
\be
(1+z^2)G(z)^2-2zG(z)+1=0.
\ee
The two solutions read
\be
G(z)=\frac{2z\pm \sqrt{4z^2-4(1+z^2)}}{2(1+z^2)}=\frac{z\pm \mathrm{i}}{1+z^2}.
\ee
Using \eqref{cauchysok}, the density comes out as expected
\be
\frac{1}{\pi}\lim_{\varepsilon\to 0}\text{Im }G(x+\mathrm{i}\varepsilon) = \frac{1}{\pi}\frac{1}{1+x^2} = \rho^\star_C(x) \;.
\ee
\underline{\textit{Constrained case.}} Now, we consider the full index problem, i.e. with an extra term in the potential as in \eqref{eq:cauchy},

\be
\frac{x}{1+x^2} + A_2\delta(x-\zeta) = \mathcal{P}\int_{-\infty}^\infty \frac{\rho^\star(y)}{x-y}dy,
\ee
where the constant $A_2$ will be determined by the new normalization condition of the rightmost blob $\int_\zeta^\infty \rho^\star(x)dx = \kappa$. We repeat the same steps as for the unconstrained integral equation, multiplying \eqref{eq:cauchy} (without the principal value) by $\frac{\rho^\star(x)}{z-x}$ and integrating over $x$. We get an extra term compared to Eq. \eqref{noprincipal}, arising from the Lagrange multiplier 
\be
\int \frac{x}{1+x^2}\frac{\rho^\star(x)}{z-x} dx = \iint\frac{\rho^\star(x)\rho^\star(y)}{(x-y)(z-y)}dxdy - \frac{A_2}{z-\zeta}.
\ee
We absorb this new term into the RHS and proceed to express, as before, all integrals in terms of $G(z)$. Our new algebraic equation will then be:
\be
 -a_1 + z(a_0-a_2) + zG(z) = (1+z^2)\left(\frac{G(z)^2}{2} -\frac{A_2}{z-\zeta}\right).
\ee
Imposing the condition that $G(z)\sim 1/z$ for $|z|\to\infty$, we get the two conditions $a_1=1/2$ and $a_0-a_2+A_2=0$. Calling $A_2=B/2$, for $\zeta=0$ we get the equation
\be
\frac{B}{2} z+G(z) z-\frac{1}{2}=\left(z^2+1\right) \left(\frac{B/2}{z }+\frac{G(z)^2}{2}\right),
\ee
whose solutions are 
\be
G(z)=\frac{z^2\pm\sqrt{-B z^3-B z-z^2}}{z^3+z} \;.
\ee
Using \eqref{cauchysok}, it is then easy to derive that the constrained density is:
\be
\rho^\star(x)=\frac{1}{\pi}\frac{\sqrt{B(x^3+x)+x^2}}{|x^3+x|},
\label{rho_with_B}
\ee
or equivalently 

\be
\rho^\star(x)=\frac{1}{\pi (1+x^2)}\sqrt{\frac{B(x+\ell_1)(x+\ell_2)}{x}}\label{rhostar} \;,
\ee
where the edge points of the leftmost blob (for $\kappa>1/2$) $-\ell_1,-\ell_2$ are determined as a function of~$B$
\begin{align}
\ell_1 &=\frac{1}{2B}(1+\sqrt{1-4 B^2}) \;, \\
\ell_2 &=\frac{1}{2B}(1-\sqrt{1-4 B^2}) .
\end{align}
The functional form in Eq. \eqref{rhostar} holds for $x$ belonging to any of the two supports.
The constant $B$ is then determined as a function of $\kappa$ by
\be
\int_0^\infty dx \frac{1}{\pi (1+x^2)}\sqrt{\frac{B(x+\ell_1)(x+\ell_2)}{x}}=\kappa \label{intconstraint} \;.
\ee
For $\kappa\to 1/2$ (unconstrained case), the solution is $B\to 0$, and $\rho^\star(x)\to \rho^\star_C(x)$ as expected.
This means that we recover the unconstrained Cauchy case if we impose a number of positive eigenvalues exactly equal to $N/2$, and this unconstrained case materializes when $B\to 0$. As $\kappa$ approaches $1/2$, the $\ell_2$ edge moves towards the origin, while the other edge $\ell_1$ approaches infinity, until exactly at the critical point $\kappa=1/2$ the two blobs merge back together. 
In Fig. \ref{fig:MC_with_function}, we show a plot of the density \eqref{rhostar} for $\kappa=0.9$. We also show standard Monte Carlo simulations of $N=300$ particles distributed according to the Boltzmann weight $\propto e^{-\beta H[{\bm \lambda}]}$ under the Hamiltonian $H[{\bm \lambda}]$ in  \eqref{hamiltonian}, with the constraint that $90\%$ of them are forced to stay on the positive semi axis . We observe a nice agreement between our exact formula and the numerics.

Finally, from \eqref{functional}, we obtain the decay of the probability of the index for large $N$ as
\be
\mathcal{P}_\beta^{(C)}(N_+,N)\approx \exp\left[-\beta N^2 \psi_C(N_+/N)\right]\label{largeNP} \;,
\ee
with
\be
\psi_C(\kappa)=\frac{1}{2}\left(S[\rho^\star]-S[\rho^\star]\Big|_{\kappa=1/2}\right)\label{rate1} \;,
\ee
where the additional term $S[\rho^\star]\Big|_{\kappa=1/2}$ comes from normalization.

In the next section, we simplify the action \eqref{eq:hamilt_1} at the saddle point and express it in terms of two single integrals which are hard to evaluate in closed form. The resulting rate function \eqref{rate1} can anyway be efficiently evaluated numerically with arbitrary precision.


\section{Computation of the rate function}
\label{computation_of_rate}

The action \eqref{eq:hamilt_1} evaluated at the saddle point density \eqref{rhostar} reads
\begin{align}
\nonumber S[\rho^\star]=&\intinf dx\rho^\star(x)\ln(1+x^2) - \int\!\!\!\intinf dxdx' \rho^\star(x)\rho^\star(x')\ln|x-x'| \\
&+ A_1\left(\intinf dx\rho^\star(x)-1\right)+A_2\left(\int_0^{+\infty}dx \rho^\star(x)-\kappa\right) \label{eq:null_terms}.
\end{align}
Since by construction $\rho^\star(x)$ satisfies the normalization conditions, the terms in the second line are automatically zero. We can now replace the double integral with single integrals. We use equation \eqref{intlog} for $\zeta=0$,
\be
2\int_{-\infty}^{+\infty}dx' \rho^\star(x')\ln|x-x'|=\ln(1+x^2) +A_1+A_2\theta(x).
\label{eq:min_hamilt_3}
\ee
Multiplying this expression by $\rho^\star(x)$ and integrating over $x$ we obtain
\be
\int\!\!\!\intinf dxdx' \rho^\star(x)\rho^\star(x')\ln|x-x'| = \frac{1}{2}\left(\intinf dx \rho^\star(x)\ln(1+x^2) +A_1+A_2\kappa\right).\label{doubleint}
\ee
Inserting \eqref{doubleint} in \eqref{eq:null_terms} we obtain
\be
S[\rho^\star]=\frac{1}{2}\intinf dx\rho^\star(x)\ln(1+x^2) -\frac{A_1}{2}-\frac{A_2}{2}\kappa.
\ee

We must now determine the constants $A_1$ and $A_2$. To do so, we first consider, without loss of generality, the case $\kappa>\frac{1}{2}$, where the density has the shape as in Fig. \ref{fig:MC_with_function}. The left blob has a compact support $[-\ell_1,-\ell_2]$. To determine the relation between $A_1$ and $A_2$, we study the asymptotic behavior of equation \eqref{eq:min_hamilt_3} when $x\to +\infty$. Since both $2\int_{-\infty}^{+\infty}dx' \rho^\star(x')\ln|x-x'|$ and $\ln(1+x^2)$ behave like $2\ln (x)$ in the large $x$ limit, we deduce that $A_1=-A_2$. Evaluating \eqref{eq:min_hamilt_3} at $x=-\ell_1$, we find the value of $A_1$, and hence $A_2$, in terms of $\ell_1$:
\be
A_1 = 2\intinf dx \rho^\star(x)\ln |x+\ell_1| - \ln(1+\ell_1^2).
\ee

\begin{figure}[!htbp]
\centering
{
	\setlength{\fboxsep}{2pt}
	\fbox{\includegraphics[width=0.8\textwidth]{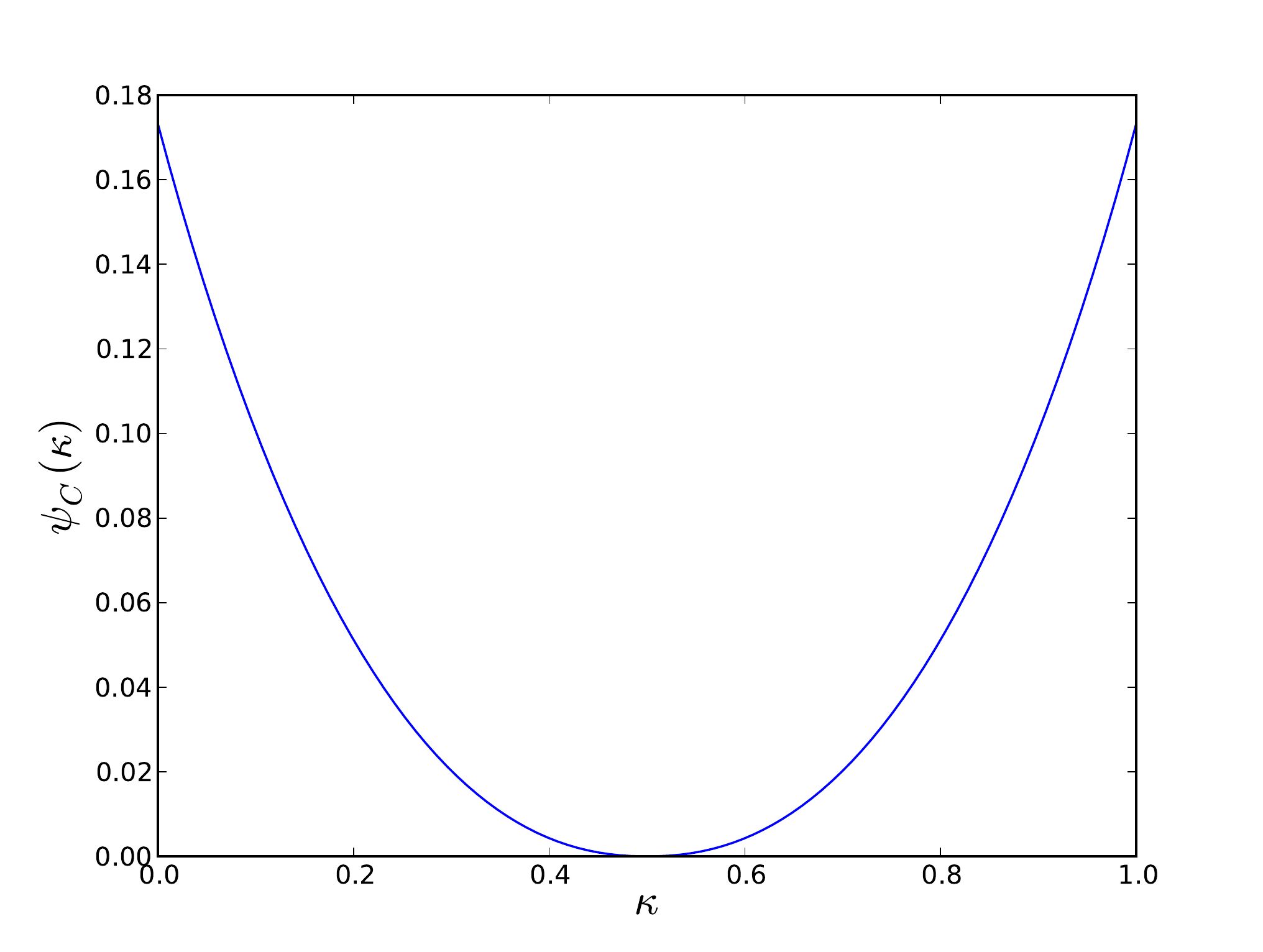}}
}
\caption{Plot of the rate function $\psi_C(\kappa)$.}
\label{fig:cauchy_action}
\end{figure}

We may finally write the action at the saddle point as
\be
S[\rho^\star]=\underbrace{\frac{1}{2}\intinf dx\rho^\star(x)\ln(1+x^2)}_{I_1}- (1-\kappa)\underbrace{\intinf dx\rho^\star(x)\ln |x+\ell_1|}_{I_2} + \frac{(1-\kappa)}{2}\underbrace{\ln(1+\ell_1^2)}_{I_3}.\label{actionI}
\ee
where $\rho^\star(x)$ (depending parametrically on $\kappa$) is given in Eq. \eqref{rhostar}. The single integrals $I_1$ and $I_2$ do not seem expressible in closed form. However the rate function \eqref{rate1} can be plotted without difficulty (see Fig. \ref{fig:cauchy_action}). One can see that the rate function is symmetric, has a minimum at $\kappa=1/2$, corresponding to the ``typical" situation, where half of the eigenvalues are positive and half are negative. In the extreme limit $\kappa=0$ (which is the same as $\kappa=1$), we get
\be
\psi_C(\kappa=0)=\frac{1}{2}\left(S[\rho^\star]\Big|_{\kappa=0}-S[\rho^\star]\Big|_{\kappa=1/2}\right)=\frac{\ln 2}{4}\approx 0.173287...
\label{cauchy_extreme}
\ee
in agreement with the large deviation law for the largest Cauchy eigenvalue [Eq. (13) in \cite{majumdar_schehr_villamaina_vivo_13} for $w=0$].

In the next section, we perform a careful asymptotic analysis of the rate function $\psi_C(\kappa)$ around the minimum $\kappa=1/2$, which brings a quadratic behavior modulated by a logarithmic singularity. This is in turn responsible for the logarithmic growth of the variance of the index with $N$ (to leading order), and provides the correct prefactor.


\section{Asymptotic analysis of $\psi_C(\kappa)$ in the vicinity of $\kappa=1/2$}
\label{asymptotic}

We have already remarked that the typical situation $\kappa=1/2$ is realized when the constant $B\to 0$. Therefore it is necessary to expand the terms $I_1,I_2$ and $I_3$ in the action \eqref{actionI}, as well as the integral constraint \eqref{intconstraint}, for $B\to 0$. It turns out that this calculation is highly nontrivial, as several cancellations occur in the leading and next-to-leading terms of each contribution (see \ref{appendix_asympt} for details). Denoting $\kappa=1/2+\delta$ (with $\delta\to 0$), the final result reads:

\begin{align}
S[\rho^\star]\sim \ln 2 + \frac{\pi}{2}B\delta + o(\delta)\label{Sdelta} \;.
\end{align}
In \ref{appendix_asympt}, we show that the relation between $\delta$ and $B$ is (to leading order for $\delta\to 0$) 

\be
\delta \sim -\frac{B\ln |B|}{\pi} \;.
\ee
Inverting this relation, we find to leading order
\be
B \sim -\pi\frac{\delta}{\ln |\delta|} \;,
\ee
implying from \eqref{Sdelta} that
\begin{align}
S[\rho^\star]\sim \ln 2 -\frac{\pi^2}{2}\frac{\delta^2}{\ln |\delta|} + o(\delta)\label{Sdelta2} \;.
\end{align}
Therefore
\be
\psi(\kappa=1/2+\delta)=S[\rho^\star]-S[\rho^\star]\Big|_{\kappa=1/2}\sim  -\frac{\pi^2}{2}\frac{\delta^2}{\ln |\delta|} \mbox{ for }\delta\to 0 \;.
\ee
From \eqref{largeNP}, we then have (for $N_+$ close to $N/2$)
\be
\mathcal{P}_\beta^{(C)}\left(N_+=\left(\frac{1}{2}+\delta\right)N,N\right)\approx\exp\left(\frac{\beta N^2}{2}\frac{\pi^2\delta^2}{2\ln |\delta|}\right) \;.\label{Pdelta}
\ee

Restoring $\delta=(N_+-N/2)/N$ in the RHS of \eqref{Pdelta}, we obtain the Gaussian behavior announced in the introduction [Eq. \eqref{Pintrovar}]:

\be
\mathcal{P}^{(C)}_\beta \left(N_+,N\right)\approx \exp\left(-\frac{\left(N_+-N/2\right)^2}{2\left(\mathrm{Var}(N_+)\right)}\right)\mbox{ for }N_+\to N/2 \;,
\ee
with
\be
 \mathrm{Var}(N_+)\sim \frac{2}{\beta \pi^2}\ln N +\mathcal{O}(1)\;.\label{varvarc}
\ee

In the next section, we compare the asymptotic behavior of the variance with a closed expression valid for finite $N$ and $\beta=2$, finding perfect agreement between the trends.


\section{Exact derivation for the variance of $N_+$ for finite $N$ and $\beta=2$}
\label{exact_N}

\begin{figure}[!htbp]
\centering
{
	\setlength{\fboxsep}{2pt}
	\fbox{\includegraphics[width=0.7\textwidth]{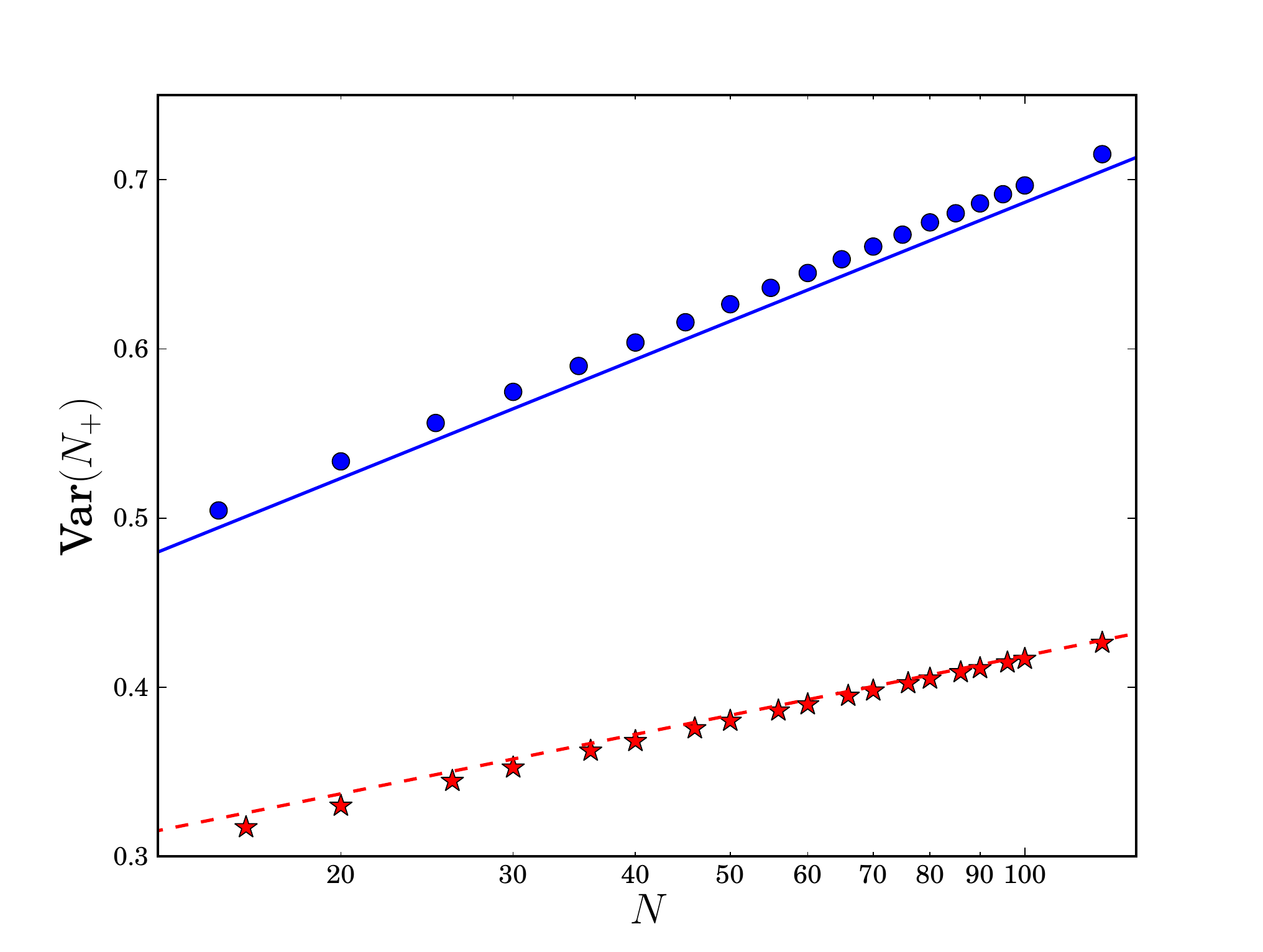}}
}
\caption{Variance of the index as a function of $N$ for $\beta=2$. Blue dots correspond to the finite $N$ exact formula \eqref{polynomials} for the Cauchy ensemble, while red stars correspond to the finite $N$ exact formula [Eq. (145) in \cite{majumdar_nadal_scardicchio_vivo_11}] for the Gaussian ensemble. The solid blue line and the dashed red line correspond to the asymptotic behaviors \eqref{varvarc} and \eqref{varlargeN} for the Cauchy and the Gaussian ensemble respectively ($\beta=2$).}
\label{fig:variance}
\end{figure}

In \ref{exact_N_der}, we derive a general formula for the variance of any linear statistics at finite $N$ and $\beta=2$. Specializing it to the index case, we deduce that
\be
\mathrm{Var}(N_+)=\frac{N}{2}-\iint_0^\infty d\lambda d\lambda^\prime [K_N(\lambda,\lambda^\prime)]^2\label{formulavartext} \;,
\ee
where $K_N(\lambda,\lambda^\prime)$ is the kernel of the ensemble, built upon suitable orthogonal polynomials. It turns out that in the Cauchy case, the orthogonal polynomials $\pi_n(x)$ satisfying
\be
\int_{-\infty}^\infty dx \frac{\pi_n(x)\pi_m(x)}{(1+x^2)^N}=\delta_{mn}
\label{ortonorm}
\ee
are
\be
\pi_n(x)=\mathrm{i}^n2^N\left[\frac{n!(N-n-\frac{1}{2})\Gamma^2(N-n)}{2\pi \Gamma(2N-n)}\right]^{1/2}P_n^{(-N,-N)}(\mathrm{i}x) \;,\label{Ijacobi} 
\ee
where $P_n^{(-N,-N)}(x)$ are Jacobi polynomials. The kernel then reads
\be
K_N(\lambda,\lambda^\prime)=\frac{1}{(1+\lambda^2)^{N/2}}\frac{1}{(1+\lambda^{\prime 2})^{N/2}}\sum_{j=0}^{N-1}\pi_j(\lambda)\pi_j(\lambda^\prime) \;,\label{kernelcauchy}
\ee
and inserting \eqref{kernelcauchy} into \eqref{formulavartext} we obtain after simple algebra
\be
\mathrm{Var}(N_+)=\frac{N}{4}-2\sum_{\stackrel{n<m}{n+m\mbox{\tiny{ odd}}}}^{N-1}\left(\int_0^\infty dx\frac{\pi_n(x)\pi_m(x)}{(1+x^2)^N} \right)^2 \;,\label{polynomials}
\ee
where we used some simple symmetry properties of those orthogonal polynomials. Unfortunately, extracting the asymptotic behavior of \eqref{polynomials} as $N\to\infty$ is not an easy task. However, formula \eqref{polynomials} can be evaluated exactly in closed form for any finite $N$ (see Eq. \eqref{finalvar} and Table \ref{table1}). The result is plotted in Fig. \ref{fig:variance} together with the corresponding result for the Gaussian ensemble, and the large $N$ logarithmic behavior in both cases. One can indeed see that the slope of the Cauchy case is twice as steep as the Gaussian case, as predicted by the asymptotic expansion of the rate function around the minimum (see Section \ref{asymptotic}). 

To simplify expression \eqref{polynomials}, we define:
\be
I_{m,n,N}=\int_0^\infty \!\!\! dx\frac{\pi_n(x)\pi_m(x)}{(1+x^2)^N}\;.\label{Ivar}
\ee
Using \eqref{Ijacobi} and the definition of Jacobi polynomials 
\be
P_n^{(-N,-N)}(\mathrm{i}x)=\sum_{k=0}^n c_{k,n}^{(N)} (1-\mathrm{i}x)^k \;,
\ee
where
\be
c_{l,\tau}^{(N)}=\frac{1}{\tau!}\frac{(-\tau)_l(-2N+\tau+1)_l(-N+l+1)_{\tau-l}}{l!2^l},
\ee
we can write

\begin{align}
\int_0^\infty  dx\frac{P_n^{(-N,-N)}(\mathrm{i}x)P_m^{(-N,-N)}(\mathrm{i}x)}{(1+x^2)^N}
=&\sum_{k=0}^m\sum_{r=0}^nc_{k,m}^{(N)}c_{r,n}^{(N)}\sum_{s=0}^{k+r}\binom{k+r}{s}(-\mathrm{i})^s\frac{1}{2}B\left(\frac{1+s}{2},N-\frac{1+s}{2}\right),
\end{align}
Here, $B(x,y)$ is Euler's Beta function. Finally, we may write the variance of index for the Cauchy ensemble as:
\be
\mathrm{Var}(N_+)=\frac{N}{4}-2\sum_{\stackrel{n<m}{n+m\mbox{\tiny{ odd}}}}^{N-1}I_{m,n,N}^2,\label{finalvar}
\ee
where
\begin{align}
\nonumber I_{m,n,N}=&\mathrm{i}^{m+n}2^{2N}\left[\frac{n!(N-n-\frac{1}{2})\Gamma^2(N-n)}{2\pi \Gamma(2N-n)}\right]^{\frac{1}{2}}\left[\frac{m!(N-m-\frac{1}{2})\Gamma^2(N-m)}{2\pi \Gamma(2N-m)}\right]^{\frac{1}{2}}\\
&\times \sum_{k=0}^m\sum_{r=0}^n\sum_{s=0}^{k+r}c_{k,m}^{(N)}c_{r,n}^{(N)}\binom{k+r}{s}(-\mathrm{i})^s\frac{1}{2}B\left(\frac{1+s}{2},N-\frac{1+s}{2}\right)
\end{align}

We here include a table with exact values of the variance for few values of $N$, together with their numerical value.

\begin{table}[h]
\centering
\begin{tabular}{|c|c|c|}
\hline $N$ & Var($N_+$) exact \\ 
\hline 2 & $\frac{1}{2} - \frac{2}{\pi^2}= 0.2973...$\\
\hline 3 & $\frac{3}{4}-\frac{4}{\pi ^2}= 0.3447...$ \\ 
\hline 5 & $\frac{5}{4}-\frac{76}{9 \pi ^2} = 0.3944...$ \\ 
\hline 10 & $\frac{5}{2}-\frac{398938}{19845 \pi ^2} = 0.4632...$ \\ 
\hline 15 & $\frac{15}{4}-\frac{4332855248}{135270135 \pi ^2} = 0.5046...$ \\ 
\hline 
\end{tabular}
\caption{Table of the variance of the Cauchy index for a few selected values of $N$.}
\label{table1}
\end{table} 

\section{Conclusion}\label{conclusion}
Using a Coulomb gas technique, originally devised by Dyson and recently employed to a variety of different situations, we compute analytically for large $N$ the probability that a $N\times N$ Cauchy matrix has a fraction $\kappa$ of eigenvalues exceeding a threshold at $\zeta$. We mainly focus on the case $\zeta=0$ for simplicity, and we find that this probability satisfies a large deviation law whose rate function $\psi_C(\kappa)$ is explicitly computed (Eqs. \eqref{rate1} and \eqref{actionI}). Expanding the rate function around its minimum $\kappa=1/2$, we find a quadratic behavior modulated by a logarithmic singularity. As a consequence, the variance of the index grows logarithmically with the matrix size $N$, with a prefactor that is twice as large as the Gaussian and Wishart ensembles \eqref{variance_cauchy}. In the limit $\kappa\to 0$ (all the eigenvalues are negative) we recover the large deviation tails of the largest Cauchy eigenvalue, recently computed in \cite{majumdar_schehr_villamaina_vivo_13}.
The logarithmic growth of the variance with $N$ is also checked against a finite $N$ formula [Eq. \eqref{finalvar}] that we derived for $\beta=2$ using orthogonal polynomials. For Cauchy random matrices, the local statistics is described by the sine-kernel \cite{BO01}, which also describes the bulk local statistics of Gaussian and Wishart random matrices. Hence the main characteristic of the Cauchy ensemble is the fact that the average spectral density extends over the full real line, as opposed to a finite support in the Gaussian and Wishart matrices. We thus expect that the absence of an edge in the case of Cauchy random  matrices is indeed responsible for this larger variance \eqref{variance_cauchy}. A more precise analysis of this effect goes beyond the scope of the present paper and is left for future investigations \cite{MMSVprep}. Other related directions of research include the determination of the subleading constant $C_\beta$ in the large $N$ expansion of the variance, based on formula \eqref{finalvar}, as well as explicit formulae for the full probability distributions for finite $N$ and all $\beta$s in the three ensembles (Gaussian, Wishart and Cauchy). 

\section*{Acknowledgments}
PV and GS acknowledge support from Labex-PALM (Project Randmat). SNM and GS acknowledge support by ANR grant 2011-BS04-013-01 WALKMAT and in part by the Indo-French Centre for the Promotion of Advanced Research under Project 4604-3.

\appendix

\section{Asymptotic analysis}
\label{appendix_asympt}

In this Appendix, we perform a careful asymptotic analysis of the rate function $\psi_C(\kappa)$ around its minimum $\kappa=1/2$. To do so, we have to estimate the behavior of $I_1,I_2$ and $I_3$, the three contributions to the action at the saddle point [Eq. \eqref{actionI}], for $B\to 0$.

First, note that the edge points $\ell_1$ and $\ell_2$ behave as $\frac{1}{B}-B$ and $B$, respectively, when $B\to 0$. Also, the support of the density (for $\kappa>1/2$) $\rho^\star(x)$ is $[-\ell_1,-\ell_2]\cup(0,+\infty)$, therefore we need to consider the two blobs of each term in the action separately.

\begin{itemize}
	\item $I_1=\frac{1}{2}\int_{-\infty}^\infty dx\rho^\star(x)\ln(1+x^2)$ \;.
\end{itemize}

First, we separate the integral as
\begin{align}
I_1=I_1^L+I_1^R & & I_1^L= \frac{1}{2}\int_{-\ell_1}^{-\ell_2}\!\!\!dx \rho^\star(x)\ln(1+x^2) & \;, & I_1^R= \frac{1}{2}\int_{0}^{+\infty}\!\!\!dx \rho^\star(x)\ln(1+x^2).
\end{align}
Writing $I_1^R$ explicitly we obtain 
\be
I_1^R= \frac{1}{2\pi}\int_{0}^{+\infty}\!\!\!dx \frac{\sqrt{B(x^3+x)+x^2}}{x(x^2+1)}\ln(1+x^2).
\label{eq:IR}
\ee
To compute the asymptotic behavior when $B\to 0$, we split this integral into two parts, one integral from 0 to $B$ and one from $B$ to $\infty$:
\be
I_1^R= \frac{1}{2\pi}\left(\int_{0}^{B}\!\!\!dx \frac{\sqrt{B(x^3+x)+x^2}}{x(x^2+1)}\ln(1+x^2)+\int_{B}^{\infty}\!\!\!dx \frac{\sqrt{B(x^3+x)+x^2}}{x(x^2+1)}\ln(1+x^2)\right).
\ee
Now we can expand the integrands in series around $B=0$ and integrate term by term to obtain [up to order ${\cal O}(B)$]
\be
I_1^R\xrightarrow[B\to 0]{} \frac{\ln 2}{2}+B\frac{\ln ^2(B)}{4\pi}-\frac{B \ln (B)}{2\pi}(1+\ln 4)+\frac{B}{4\pi}\left(2+4\ln 2(1+\ln 2)+\frac{3\pi^2}{4}\right)+o(B).
\ee

We now turn our attention to $I_1^L$, calculating the asymptotic behavior when $B\to 0$ of the integral:
\be
I_1^L= \frac{1}{2\pi}\int_{-\ell_1}^{-\ell_2}\!\!\!dx \frac{\sqrt{B(x^3+x)+x^2}}{|x(x^2+1)|}\ln(1+x^2).
\label{eq:IL_B}
\ee
To proceed, it is more convenient to reexpress $I_1^L$ in terms of its edge points
\be
I_1^L= \frac{\sqrt{B}}{2\pi}\int_{-\ell_1}^{-\ell_2}\!\!\!dx \frac{\sqrt{-(x+\ell_1)(x+\ell_2)}}{\sqrt{|x|}(x^2+1)}\ln(1+x^2),
\label{eq:IL_l1l2}
\ee
which is equivalent to \eqref{eq:IL_B}. We proceed with the following change of variable $y=\frac{x+\ell_1}{\ell_1-\ell_2}$, we have:
\be
I_1^L= \frac{\sqrt{B}}{2\pi}\int_{0}^{1}\!\frac{(\ell_1-\ell_2)dy}{1+[(\ell_1-\ell_2)y-\ell_1]^2}\frac{(\ell_1-\ell_2)\sqrt{y(1-y)}}{\sqrt{|(\ell_1-\ell_2)y-\ell_1|}}\ln\left[1+ [(\ell_1-\ell_2)y-\ell_1]^2\right].\label{I1Las}
\ee
Since $\ell_1$ behaves like $\frac{1}{B}-B$ and $\ell_2$ behaves like $B$ when $B$ is small, $\ell_1-\ell_2$ goes as $\frac{1}{B}-2B$. We replace these asymptotic behaviors in \eqref{I1Las}, keeping only the leading orders for small $B$. The resulting integral can be computed explicitly and we can then extract its asymptotic behavior when $B\to 0$

\be
I_1^L \xrightarrow[B\to 0]{}\frac{\ln 2}{2} -\frac{B \ln ^2B}{4 \pi }+\frac{(1+\ln 4) (B \ln B)}{2 \pi } + \frac{B}{4\pi}\left(\frac{\pi ^2}{4}-2-4 \ln 2 (1+\ln 2)\right)+o(B)
\ee

Note an impressive series of cancellations in the sum $I_1^L+I_1^R$, resulting in 
\be
I_1 \xrightarrow[B\to 0]{}\ln 2+ \frac{\pi}{4}B + o(B).
\ee

\begin{itemize}
	\item $ I_2=\int_{-\infty}^{\infty}dx \rho^\star(x)\ln(\ell_1+x)$.
\end{itemize}
First, we again separate the integral over the two supports
\begin{align}
I_2=I_2^L+I_2^R & & I_2^L= \int_{-\ell_1}^{-\ell_2}\!\!\!dx \rho^\star(x)\ln(\ell_1+x) & & I_2^R= \int_{0}^{+\infty}\!\!\!dx \rho^\star(x)\ln(\ell_1+x).
\end{align}
Both integrals are very similar to $I_1^R$ and $I_R^L$, and we proceed to calculate them by the same methods. Expanding the integrals to leading orders of $B$ and adding both terms, we get to

\be
I_2 \xrightarrow[B\to 0]{}-\ln B+ \frac{\pi}{2}B + o(B).
\ee
The third term, $I_3$, has a straightforward expansion
\be
I_3 \xrightarrow[B\to 0]{} -2\ln B +  o(B).
\ee
The action $S[\rho^\star]$ at the saddle point has therefore an expansion for $B\to 0$ equal to
\begin{align}
S[\rho^\star]\sim \ln 2 + \frac{\pi}{4}B - (1-\kappa)\left(-\ln B+\frac{\pi}{2}B\right)+\frac{(1-\kappa)}{2}(-2\ln B)+o(B). \label{S_with_B}
\end{align}
Once we write $\kappa=\frac{1}{2}+\delta$, we obtain Eq. \eqref{Sdelta}.

\section{General formula for the variance of a linear statistics at finite $N$ for $\beta=2$}
\label{exact_N_der}

We derive here a general fluctuation formula (valid for $\beta=2$ and finite $N$) for the variance of a \emph{linear statistics}, i.e. a random variable $\phi$ of the form
\be
\phi=\sum_{i=1}^N f(\lambda_i) \;,
\ee
for a general benign function $f(x)$. The variance of $\phi$ is given by $\mathrm{Var}(\phi)=\langle\phi^2\rangle-\langle\phi\rangle^2$, where the average is taken with respect to the jpd of the $N$ eigenvalues $P(\lambda_1,\ldots,\lambda_N)$.

By definition
\be
\langle\phi^2\rangle=\idotsint d\lambda_1\ldots d\lambda_N P(\lambda_1,\ldots,\lambda_N)\sum_{i=1}^N f(\lambda_i)\sum_{j=1}^N f(\lambda_j).
\label{linear_stat}
\ee

Let us first define the one-point and the two-point correlation function (marginals) of the jpdf $P(\lambda_1,\ldots,\lambda_N)$\footnote{The one-point function $R_1(\lambda)$ is related to the finite-$N$ spectral density via $R_1(\lambda)=N\rho_N(\lambda)$.}
\begin{align}
R_1(\lambda) &=N\idotsint d\lambda_2\ldots d\lambda_N P(\lambda,\lambda_2,\ldots,\lambda_N),\\
R_2(\lambda,\lambda')=&N(N-1) \idotsint d\lambda_3\ldots d\lambda_N P(\lambda,\lambda',\lambda_3,\ldots,\lambda_N)\;.\label{R2}
\end{align}
Separating in the double sum in \eqref{linear_stat} the term $i=j$ from the terms $i\neq j$ we easily obtain
\be
\langle\phi^2\rangle=\int d\lambda R_1(\lambda)\left[ f(\lambda) \right]^2+\iint d\lambda d\lambda' f(\lambda)f(\lambda')R_2(\lambda,\lambda')\label{squarephi} \;,
\ee
while clearly
\be
\langle\phi\rangle^2=\left[\int d\lambda R_1(\lambda)f(\lambda)\right]^2 \;.
\ee
The theory of orthogonal polynomials \cite{mehta} gives a prescription to compute $n$-point correlation functions for $\beta=2$ in terms of a $n\times n$ determinant. More precisely, let

\be
P(\lambda_1,\ldots,\lambda_N)\propto\prod_{j<k}|\lambda_j-\lambda_k|^2 \prod_{j=1}^N e^{-V(\lambda_j)}\label{jpdgeneralbeta2} \;.
\ee
Then the associated \emph{kernel} is
\be
K_N(\lambda,\lambda^\prime) = e^{-\frac{1}{2}(V(\lambda)+V(\lambda^\prime))}\sum_{j=0}^{N-1}\pi_j(\lambda)\pi_j(\lambda^\prime) \;,
\ee
where the $\pi_j$ are orthogonal polynomials with respect to the weight $e^{-V(\lambda)}$, i.e.
\be
\int d\lambda e^{-V(\lambda)}\pi_m(\lambda)\pi_n(\lambda)=\delta_{mn} \;.
\ee
Then
\begin{align}
R_1(\lambda) &= K_N(\lambda,\lambda)\label{R1al} \;, \\
R_2(\lambda,\lambda^\prime) &=K_N(\lambda,\lambda)K_N(\lambda^\prime,\lambda^\prime)-(K_N(\lambda,\lambda^\prime))^2\label{R2al} \;.
\end{align}
Inserting \eqref{R1al} and \eqref{R2al} into \eqref{squarephi}, and using $R_1(\lambda)=K_N(\lambda,\lambda)=N\rho_N(\lambda)$ we eventually get
\be
\mathrm{Var}(\phi)=N\int d\lambda \rho_N(\lambda)[f(\lambda)]^2-\iint d\lambda d\lambda^\prime f(\lambda)f(\lambda^\prime)[K_N(\lambda,\lambda^\prime)]^2 \;.
\ee
Specializing this formula to the index case, where $f(\lambda)=\theta(\lambda)$ we get
\be
\mathrm{Var}(N_+)=\frac{N}{2}-\iint_0^\infty d\lambda d\lambda^\prime [K_N(\lambda,\lambda^\prime)]^2 \;.
\ee
as in Eq. \eqref{formulavartext}.

\end{document}